\documentclass[aps, prl, preprint, groupedaddress, superscriptaddress]{revtex4-1}

\usepackage{graphicx}   
\usepackage{amsmath}   
\usepackage{siunitx}
\usepackage[utf8]{inputenc}

\begin{document}

\title{Direct imaging of antiferromagnetic domains in Mn$_2$Au manipulated by high magnetic fields}

\author{A.A.\,Sapozhnik}
\affiliation{Institut f\"ur Physik, JGU Mainz, Staudingerweg 7, 55128 Mainz, Germany}
\affiliation{Graduate School Materials Science in Mainz, Staudingerweg 9, 55128 Mainz, Germany}
\author{M.\,Filianina}
\affiliation{Institut f\"ur Physik, JGU Mainz, Staudingerweg 7, 55128 Mainz, Germany}
\affiliation{Graduate School Materials Science in Mainz, Staudingerweg 9, 55128 Mainz, Germany}
\author{S.Yu.\,Bodnar}
\affiliation{Institut f\"ur Physik, JGU Mainz, Staudingerweg 7, 55128 Mainz, Germany}
\author{A.\,Lamirand}
\affiliation{Diamond Light Source, Chilton, Didcot, Oxfordshire, OX11 0DE, UK}
\author{M.\,Mawass}
\affiliation{Helmholtz-Zentrum Berlin f\"ur Materialien und Energie, Albert-Einstein Str. 15, 12489, Berlin, Germany}
\author{Y.\,Skourski}
\affiliation{Hochfeld-Magnetlabor Dresden (HLD-EMFL), Helmholtz-Zentrum Dresden-Rossendorf, 01328 Dresden, Germany}
\author{H.-J.\,Elmers}
\affiliation{Institut f\"ur Physik, JGU Mainz, Staudingerweg 7, 55128 Mainz, Germany}
\affiliation{Graduate School Materials Science in Mainz, Staudingerweg 9, 55128 Mainz, Germany}
\author{H.\,Zabel}
\affiliation{Institut f\"ur Physik, JGU Mainz, Staudingerweg 7, 55128 Mainz, Germany}
\affiliation{Graduate School Materials Science in Mainz, Staudingerweg 9, 55128 Mainz, Germany}
\author{M.\,Kläui}
\affiliation{Institut f\"ur Physik, JGU Mainz, Staudingerweg 7, 55128 Mainz, Germany}
\affiliation{Graduate School Materials Science in Mainz, Staudingerweg 9, 55128 Mainz, Germany}
\author{M.\,Jourdan}
\affiliation{Institut f\"ur Physik, JGU Mainz, Staudingerweg 7, 55128 Mainz, Germany}
\affiliation{Graduate School Materials Science in Mainz, Staudingerweg 9, 55128 Mainz, Germany}

\begin{abstract}
In the field of antiferromagnetic (AFM) spintronics, information about the Néel vector, AFM domain sizes, and spin-flop fields is a prerequisite for device applications but is not available easily. We have investigated AFM domains and spin-flop induced changes of domain patterns in Mn$_2$Au(001) epitaxial thin films by X-ray magnetic linear dichroism photoemission electron microscopy (XMLD-PEEM) using magnetic fields up to \SI{70}{T}. As-prepared Mn$_2$Au films exhibit AFM domains with an average size $\leq$\SI{1}{\micro\metre}. Application of a \SI{30}{\tesla} field, exceeding the spin-flop field, along a magnetocrystalline easy axis dramatically increases the AFM domain size with Néel vectors perpendicular to the applied field direction. The width of Néel type domain walls (DW) is below the spatial resolution of the PEEM and therefore can only be estimated from an analysis of the DW profile to be smaller than \SI{80}{\nano\metre}. Furthermore, using the values for the DW width and the spin-flop field, we evaluate an in-plane anisotropy constant ranging between 1 and \SI{17}{\micro\electronvolt}/f.u.. 
\end{abstract}


\maketitle

\section{I. INTRODUCTION}

In antiferromagnetic (AFM) spintronics, ferromagnets (FM) are replaced by AFMs as active device materials \cite{MAC11, GOM17, GOM14, JUN16, BAL17}. This novel approach takes advantage of the fast THz dynamics of antiferromagnets, which is driven by exchange interaction. This in principle enables writing speeds superior to those in conventional spintronics based on FM. Additionally, the absence of a net magnetization in AFMs results in vanishing dipolar interactions allowing for an increased information density and a high stability against disturbing external fields.

In AFM spintronics, information is encoded by the direction of the Néel vector, which is defined by the vectorial difference of the sublattice magnetizations. For spintronic applications, one requires efficient methods for reading and changing the Néel vector orientation (writing). In case of AFMs with non-centrosymmetric magnetic sublattices, it was predicted that a current induced spin-orbit torque can change the orientation of the Néel vector \cite{ZEL14}. This was indeed recently demonstrated for the compounds CuMnAs \cite{WAD15, OLE17} and Mn$_2$Au \cite{BOD18, MEI17}. However, the necessary current densities are in general close to the destruction limit and further materials optimization is required, specifically aiming at a reduced DW pinning. To this end, the characterization of the AFM domain structure is of major importance. Whereas X-ray magnetic linear dichroism photoemission electron microscopy (XMLD-PEEM) was successfully used to observe AFM domains and their manipulation in CuMnAs thin films \cite{WAD15, WAD17}, no such experiments have been reported for Mn$_2$Au yet. Moreover, information about the magnetocrystalline anisotropy determining switching current \cite{ZEL14} is required for lowering its threshold density. 

Magnetic domains in different AFMs were widely studied by XMLD-PEEM during the last two decades \cite{STO99, NOL00, KRU08, WAD15, BALD17}. The common procedure developed for obtaining XMLD-contrast in oxides is to calculate the asymmetry from images taken at two energies corresponding to multiplet peaks of a magnetic atom. Typical domain sizes observed in oxide AFM are in the micrometer regime \cite{STO99, NOL00, KRU08}. However, conductive materials, like Mn$_2$Au, exhibit broader X-ray absorption spectra (XAS) with no multiplet structure \cite{SAP17}. This renders the selection of the appropriate X-ray energies for obtaining sufficient magnetic contrast challenging \cite{GRZ17}.

In order to understand Mn$_2$Au and use this novel material for future devices, one needs to be able to visualize the magnetic domain configuration and obtain key magnetic properties, such as the anisotropies, which are currently unknown. We report on the visualization of AFM domains in Mn$_2$Au epitaxial thin films using high resolution XMLD-PEEM. We demonstrate the alignment of the Néel vector by the application of a high magnetic field resulting in a spin-flop transition. This allows us based on an analysis of the AFM domain wall width and on the spin-flop field to evaluate the magnetic in-plane anisotropy constant of Mn$_2$Au.   

\section{II. METHODS}

Epitaxial thin film samples with a stacking sequence of Al$_2$O$_3$\,(1$\bar{1}$02)\, substrate/Ta(001)\,\SI{30}{\nano\metre}/\\/Mn$_2$Au(001)\,\SI{240}{\nano\metre}/AlO$_x$\,\SI{2}{\nano\metre} were grown by radio frequency magnetron sputtering. The in-plane epitaxial relation as determined by X-ray diffractometry (XRD) is Ta(001)[100] $\parallel$ Mn$_2$Au(001)[100]. AlO$_x$ capping was used as an oxidation protection. More details on the sample preparation and characterization can be found in Ref.\,\cite{JOU15}.

The samples were exposed to high pulsed magnetic fields targeting an alignment of the Néel vector by a spin-flop transition.  Magnetic fields of different amplitudes ranging from \SI{30}{\tesla} to \SI{70}{\tesla} were applied to the Mn$_2$Au samples along the [110] and [100]-directions at room temperature at the High Magnetic Field Laboratory of the Helmholtz-Zentrum Dresden-Rossendorf (HZDR).

The XMLD-PEEM studies were performed at beamline I06 at Diamond Light Source and with the SPEEM setup at BESSY\,II (HZB). Both instruments are equipped with Elmitec photoemission electron microscopes providing $\sim$\SI{50}{\nano\metre} spatial resolution and \SI{0.4}{\electronvolt} energy resolution. The X-ray beam linearly polarized in the sample plane was incident under an angle of 16$^\circ$ to the sample surface. Controllable rotation procedures guaranteed measurements within the same area on the surface at different rotation angles around the sample normal. The strongest in-plane magnetic contrast was achieved by calculating the asymmetry of two images taken at energies corresponding to the maximum ($E_{MAX}$) and to the minimum ($E_{MIN}$) of XMLD:

\begin{equation}
	I_{asym} = \frac{I(E_{MAX}) - I(E_{MIN})}{I(E_{MAX}) + I(E_{MIN})}.
\end{equation}

The XAS was determined from a set of images obtained in a range of energies close to the L$_3$ absorption edge of Mn. The absorption coefficient was calculated as the sum of gray-scale levels over a region of interest in the center of the field of view. In our previous work we measured the absorption spectrum of Mn$_2$Au \cite{SAP17} and demonstrated that $E_{MAX}$ and $E_{MIN}$ are separated by \SI{0.8}{\electronvolt} and \SI{0.0}{\electronvolt} from the $L_3$ absorption edge, respectively. This information, in combination with the XAS determined from the images as discussed above, was used for defining $E_{MAX}$ and $E_{MIN}$ for each sample.

\section{III. EXPERIMENTAL RESULTS AND DISCUSSION}

\begin{figure}[ht]
\includegraphics[width=\linewidth]{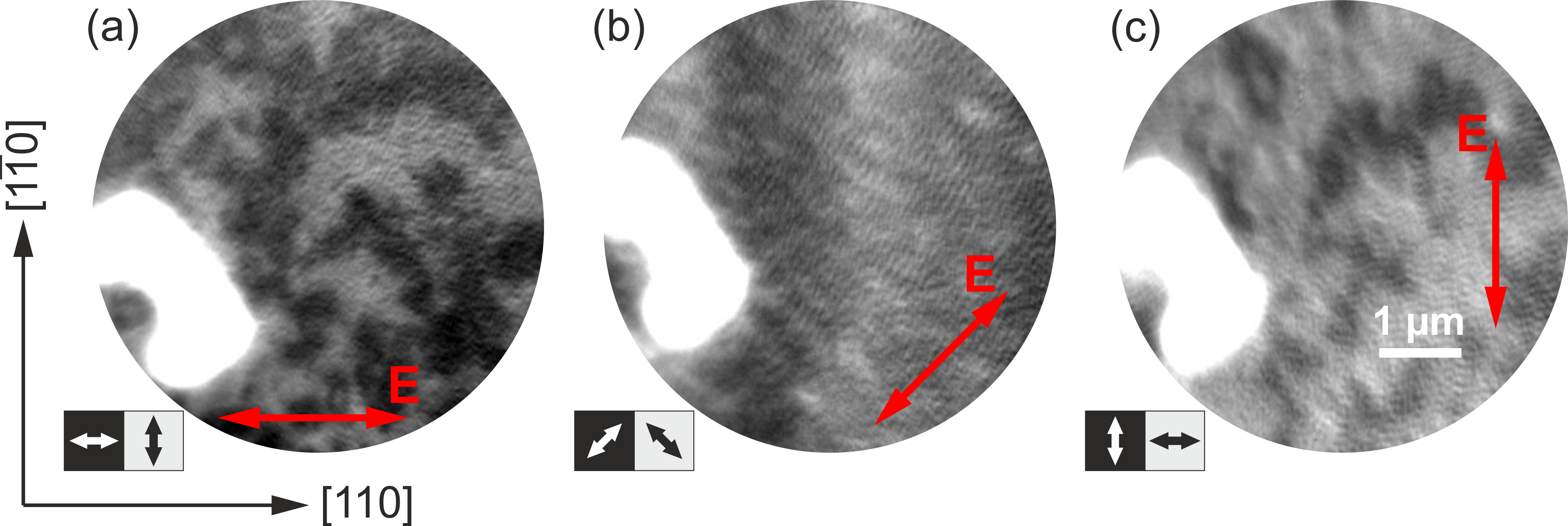}
\caption{Asymmetry images of the as-prepared sample. The in-plane angle of the X-ray incidence is (a) 0$^\circ$, (b) 45$^\circ$, and (c) 90$^\circ$ with respect to the crystallographic [1$\mathrm{\bar{1}}$0]-axis. The red double-headed arrow indicates electric field vector E of the linearly polarized X-ray beam. The double box at the bottom specifies the Néel vector orientation in the AFM domains. The bright area on the left hand side of the image is caused by the marker, which was used to keep the image position during rotation of the sample.}
\label{fig:AsPrep} 
\end{figure}

The as-prepared sample not exposed to a high magnetic field exhibits small contrast features with an average size of $\sim$\SI{1}{\micro\metre} in the asymmetry image (Fig.\,\ref{fig:AsPrep} (a)). The contrast disappears when the sample (corresponding to the direction of X-ray incidence) is rotated by 45$^\circ$ (Fig.\,\ref{fig:AsPrep} (b)) and reverses after 90$^\circ$ rotation of the sample (Fig.\,\ref{fig:AsPrep} (c)), which demonstrates the magnetic origin of the observed asymmetry. From the vanishing contrast of Fig.\,\ref{fig:AsPrep} (b) and the appearance of basically two levels of gray in Fig.\,\ref{fig:AsPrep} (a) and (c), we conclude that the Néel vector is always oriented parallel to the $\langle$110$\rangle$-directions, which is consistent with the reported easy axes of Mn$_2$Au \cite{SHI10, BAR15}. Thus, an as-prepared Mn$_2$Au sample shows AFM domain pattern as revealed in Fig.\,\ref{fig:AsPrep} with an average domain size of $\sim$\SI{1}{\micro\metre}. Moreover, the coverage of the sample with the two energetically equivalent domains with 90$^\circ$ different Néel vector orientations are comparable.

\begin{figure}[ht]
\includegraphics[width=\linewidth]{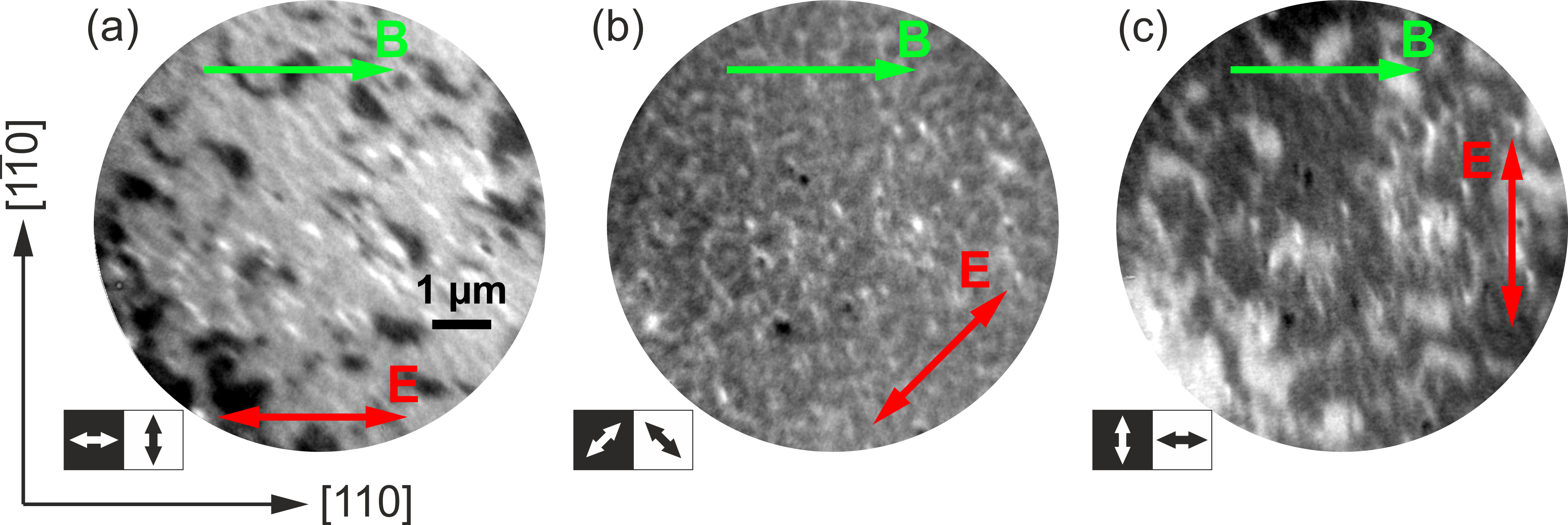}
\caption{Asymmetry images of the Mn$_2$Au sample after exposure to a magnetic field of \SI{30}{\tesla} along the [110]-direction (green arrow). The in-plane angle of the X-ray incidence is (a) 0$^\circ$, (b) 45$^\circ$, and (c) 90$^\circ$. The red double-headed arrow indicates the polarization of the linearly polarized X-ray beam. The double box at the bottom specifies the Néel vector orientation in the AFM domains.}
\label{30T_EA} 
\end{figure}

Having established the as-grown domain structure, the next step is to study changes in the domain structure when the AFM is exposed to magnetic fields sufficiently high to potentially manipulate the AFM. Fig.\,\ref{30T_EA} shows XMLD-PEEM images of a sample exposed to a \SI{30}{\tesla} external field along the [110]-direction, which is an easy-axis of the material. A strong magnetic contrast in the asymmetry image appears for the X-ray incidence direction (surface projected) parallel to [1$\mathrm{\bar{1}}$0] (0$^\circ$), which displays large bright (light gray) areas with minor dark inclusions (Fig.\,\ref{30T_EA} (a)). Again, the contrast reverses upon rotation of the sample by 90$^\circ$ with respect to the direction of X-ray incidence (Fig.\,\ref{30T_EA} (c)) demonstrating the magnetic origin of the asymmetry. Please note that the magnetic contrast vanishes upon rotation by 45$^\circ$ leaving only some morphology related features visible (Fig.\,\ref{30T_EA} (b)). A field of \SI{30}{\tesla} significantly increases the size of domains with the Néel vector perpendicular to the field. This phenomenon can be explained by a spin-flop transition, which results in reorientation of the Néel vector perpendicular to the direction in which the magnetic field was applied. From the image we see, that the majority of the AFM spin structure has been aligned with the axis favored by the spin-flop. So from this we can deduce an upper bound for the spin-flop field of our Mn$_2$Au thin films of \SI{30}{\tesla}.

\begin{figure}[ht]
\includegraphics[width=\linewidth]{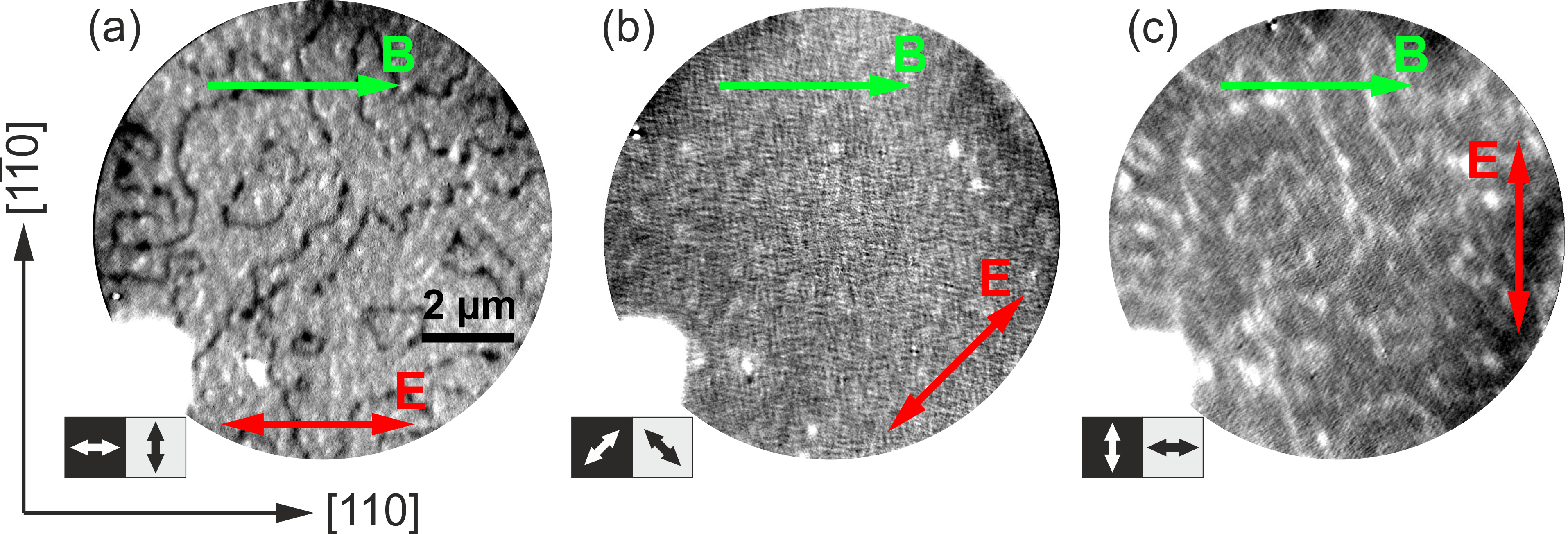}
\caption{Asymmetry images of the Mn$_2$Au sample after exposure to a magnetic field of \SI{50}{\tesla} along the [110]-direction (green arrow). The in-plane angle of the X-ray incidence is (a) 0$^\circ$, (b) 45$^\circ$, and (c) 90$^\circ$. The red double-headed arrow indicates the polarization of the linearly polarized X-ray beam. The double box at the bottom specifies the Néel vector orientation in the AFM domains.}
\label{50T_EA} 
\end{figure}

A higher magnetic field of \SI{50}{\tesla} applied along the [110]-direction causes a similar reorientation of the domain structure (Fig.\,\ref{50T_EA}). The asymmetry image shows bright (light gray) areas separated by narrow worm-like dark lines (Fig.\,\ref{50T_EA} (a)). They have an average width of $\sim$\SI{100}{\nano\metre}, which requires a higher resolution for resolving the spin structure within them. The dark lines can either be magnetic domains with the Néel vector oriented along the [110]-direction or can be considered 180$^\circ$ domain walls. The presence of not closed lines can be seen as evidence against the domain wall hypothesis. However, the gaps in the black lines can be caused by the surface morphology contributing to the asymmetry via e.g.\,residual drifts in the images remaining even after corrections.

\begin{figure}[ht]
\includegraphics[height=6cm]{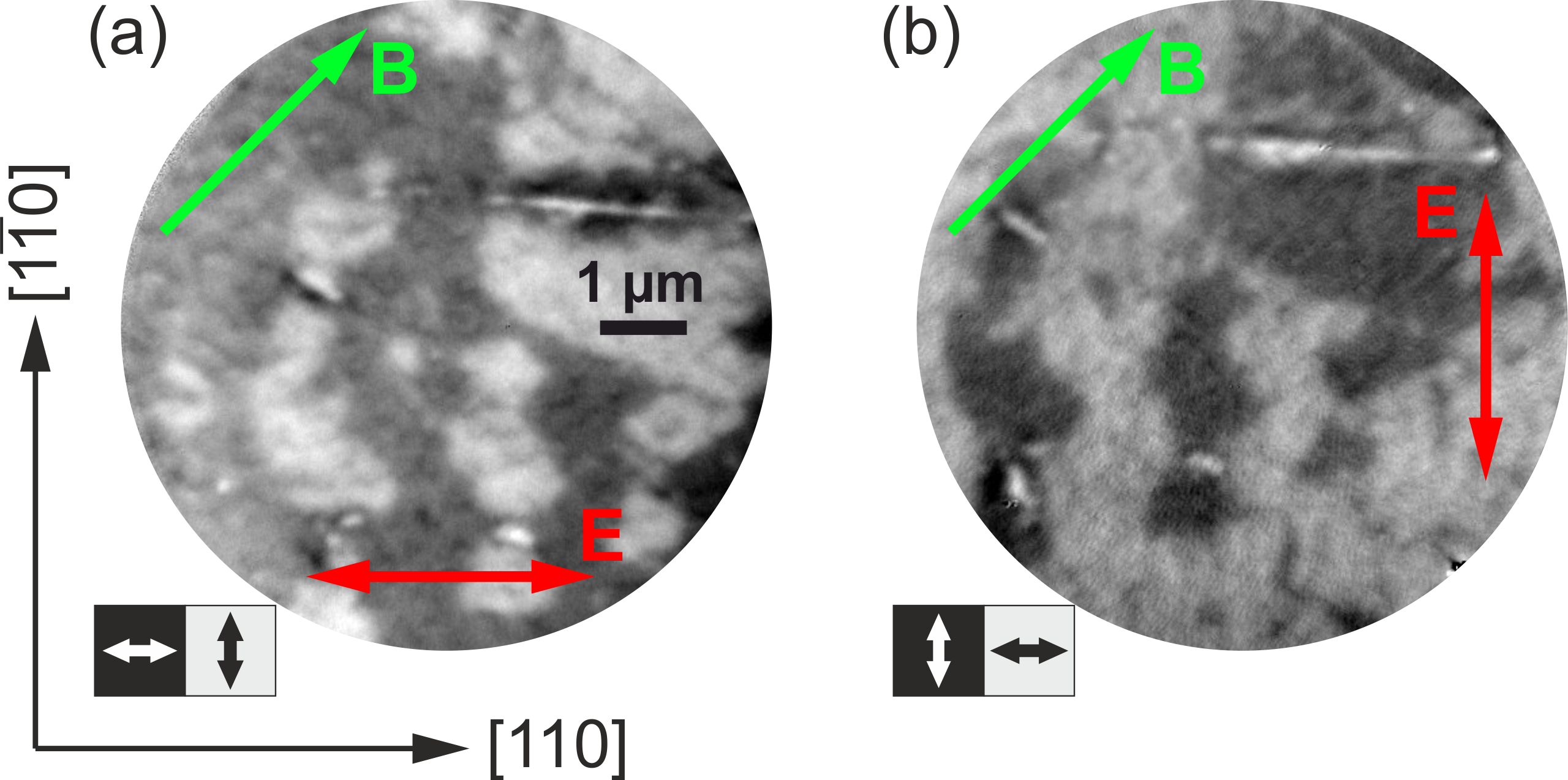}
\caption{Asymmetry images of the Mn$_2$Au sample after exposure to a magnetic field of \SI{70}{\tesla} along the [100]-direction (green arrow). The in-plane angle of the X-ray incidence is (a) 0$^\circ$, (b) 90$^\circ$. The red double-headed arrow indicates the polarization of the linearly polarized X-ray beam.}
\label{70T_HA} 
\end{figure}

Finally, we probe the effect of a \SI{70}{\tesla} external field applied along the [100]-axis of Mn$_2$Au, which is a hard magnetic direction (Fig.\,\ref{70T_HA}). The AFM domain structure is decomposed into domains with an average size of $\sim$1 to \SI{3}{\micro\metre}. However, the proportion of both types of AFM domains is equal, indicating no preferred Néel vector orientation on a large scale.

This observation is explained by the fact that a high enough external field applied along the hard [100]-axis orients the Néel vector along the perpendicular [010]-hard axis. When the field is reduced, the moments redistribute themselves parallel to the easy axes creating a new domain pattern. This leads to an increase of the average domain size in comparison to the as-prepared state shown in Fig.\,1. These results again confirm $\langle$110$\rangle$ to be the easy axes in our Mn$_2$Au thin films \cite{SHI10, BAR15}.

The upper boundary specified for the field required to generate a spin-flop transition allows us to determine the in-plane anisotropy constant ($H^{IP}_{a}$) of Mn$_2$Au(001) utilizing the following expression for spin-flop field $H_{SF} = \sqrt{2H_{ex}H^{IP}_{a}}$ \cite{MAC17}. Using the exchange field from Ref.\,\cite{BAR13} to be $\mu_0H_{exch}$=\SI{1300}{\tesla}, we find an upper boundary of $\mu_0H^{IP}_{a}$\,=\,\SI{0.35}{\tesla}. Adopting the expression for the in-plane anisotropy $K_4\,\sin^4\theta\,\cos(4\phi)$ from \cite{SHI10}, the derived anisotropy field corresponds to the maximal value of $K_{4\parallel}$\,$\leq$\,\SI{17}{\micro\electronvolt}/f.u., which is in line with theoretical predictions of \SI{10}{\micro\electronvolt}/f.u. \cite{SHI10}.

\begin{figure}[ht]
\includegraphics[width=\linewidth]{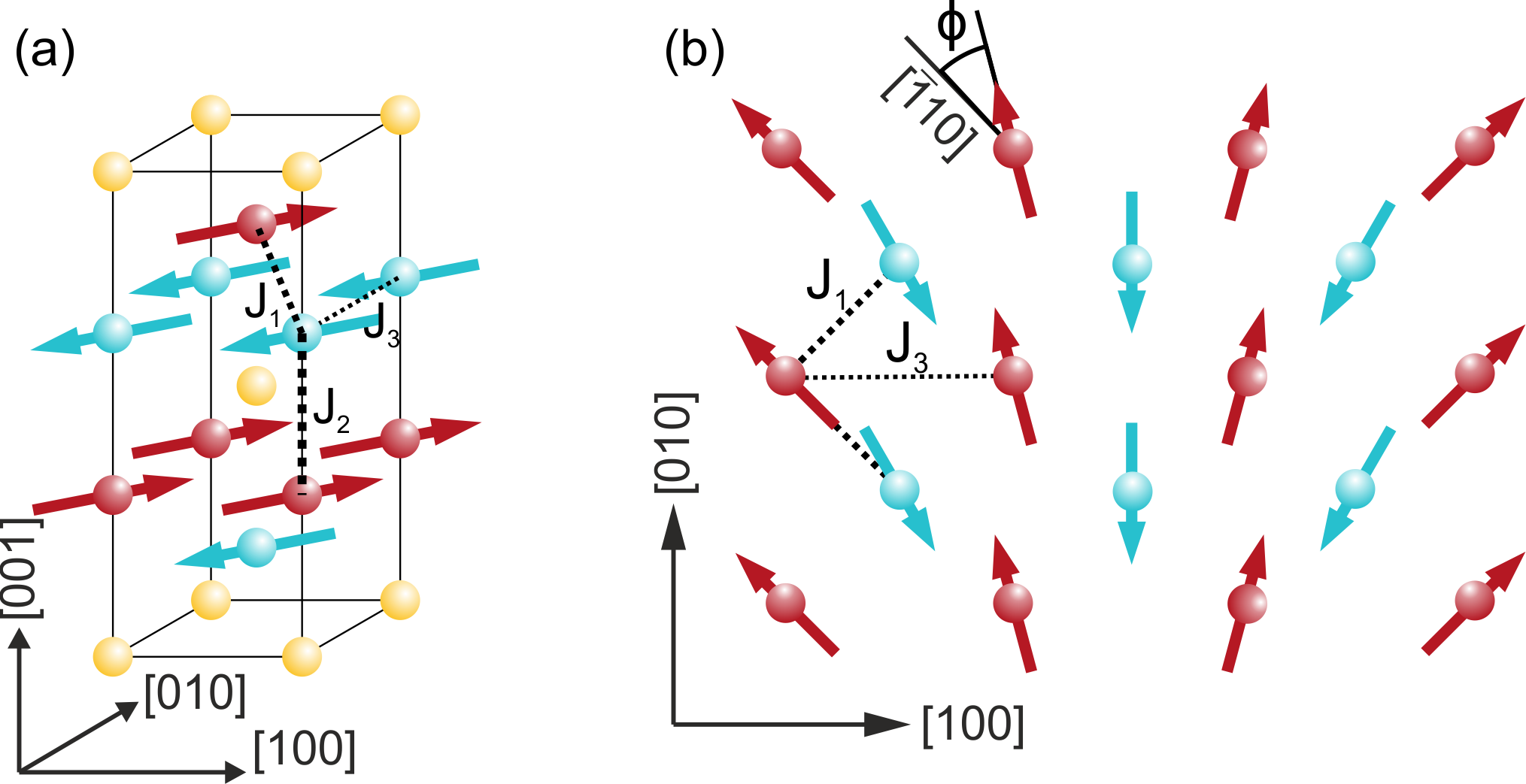}
\caption{(a) The unit cell of Mn$_2$Au indicating the exchange constants. (b) Schematic representation of a Néel type domain wall when viewed along the [001]-axis.}
\label{DW} 
\end{figure}

\section{IV. DETERMINATION OF THE DOMAIN WALL WIDTH IN $\mathbf{Mn_2Au}$}

Due to a high c-axis magnetocrystalline anisotropy in Mn$_2$Au \cite{SHI10}, in-plane (Néel type) domain walls are energetically more favorable in our thin films than Bloch walls with an out-of-plane component. The intrinsic DW width can be determined by minimizing the total DW energy per unit area, which contains exchange interaction and magnetocrystalline anisotropy terms \cite{YAM66}:

\begin{equation}
E[\phi(x)]=\int_{-\infty}^{\infty}\left(Ja^2\left(\frac{d\phi}{dx}\right)^2 + K_4(1-\cos 4\phi) \right)n\;dx,
\label{DWenergy}
\end{equation}
where $J=\frac{1}{4}(J_1+2J_3)$ (see Fig.\,\ref{DW} (a)) for [100]-domain walls (see Fig.\,\ref{DW} (b)), $K_4$ is the four-fold anisotropy constant, $n$ is the volume density of Mn atoms, and $a$ is the Mn$_2$Au lattice constant. The choice of the anisotropy term corresponds to the angle $\phi$ between the [$\mathrm{\bar{1}}$10]-easy axis and the staggered magnetization direction (Fig.\,\ref{DW} (b)). The solution of the variational problem for the functional in Eq. (\ref{DWenergy}) with the boundary conditions $\phi(-\infty)$=0 and $\phi(\infty)$=$\pi$/2 provides the DW profile:

\begin{equation}
\phi_{[100]}(x) = \arctan\;\exp\left({\sqrt{\frac{8K_4}{Ja^2}}x}\right).
\label{DWprofile2}
\end{equation}

The DW width can be expressed by the slope of $\phi(x)$ at $x$ = 0, i.\,e.\,in the center of the DW. Using Eq. (\ref{DWprofile2}), the 90$^\circ$ DW width is derived as:

\begin{equation}
w=\frac{\pi}{2}\frac{1}{d\phi_{[100]}(x)/dx(x=0)}=\frac{\pi}{2}\sqrt{\frac{J}{2K_4}}a.
\label{DWwidth}
\end{equation}
Please note that similar considerations apply for determining the width of [110]-domain walls, resulting in the same expression (\ref{DWwidth}).

Since the XMLD does not change sign upon Néel vector inversion, the normalized XMLD-PEEM contrast across a DW is proportional to the cosine squared of the angle between the Néel vector and the X-ray electric field $I_{DW}(x) \propto \cos^2\phi(x)$ \cite{ADL95}. Thus, in analogy with Eq. (\ref{DWwidth}), the antiferromagnetic DW width observed in the experiment is:

\begin{equation}
w_{exp}=\frac{1}{dI_{DW}(x)/dx(x=0)}=\sqrt{\frac{J}{2K_4}}a.
\label{DWwidthExp}
\end{equation}

This result indicates that a DW appears $\pi/2$ times more narrow in a XMLD-PEEM image in comparison with the actual width.

\begin{figure}[ht]
\includegraphics[height=3.75cm]{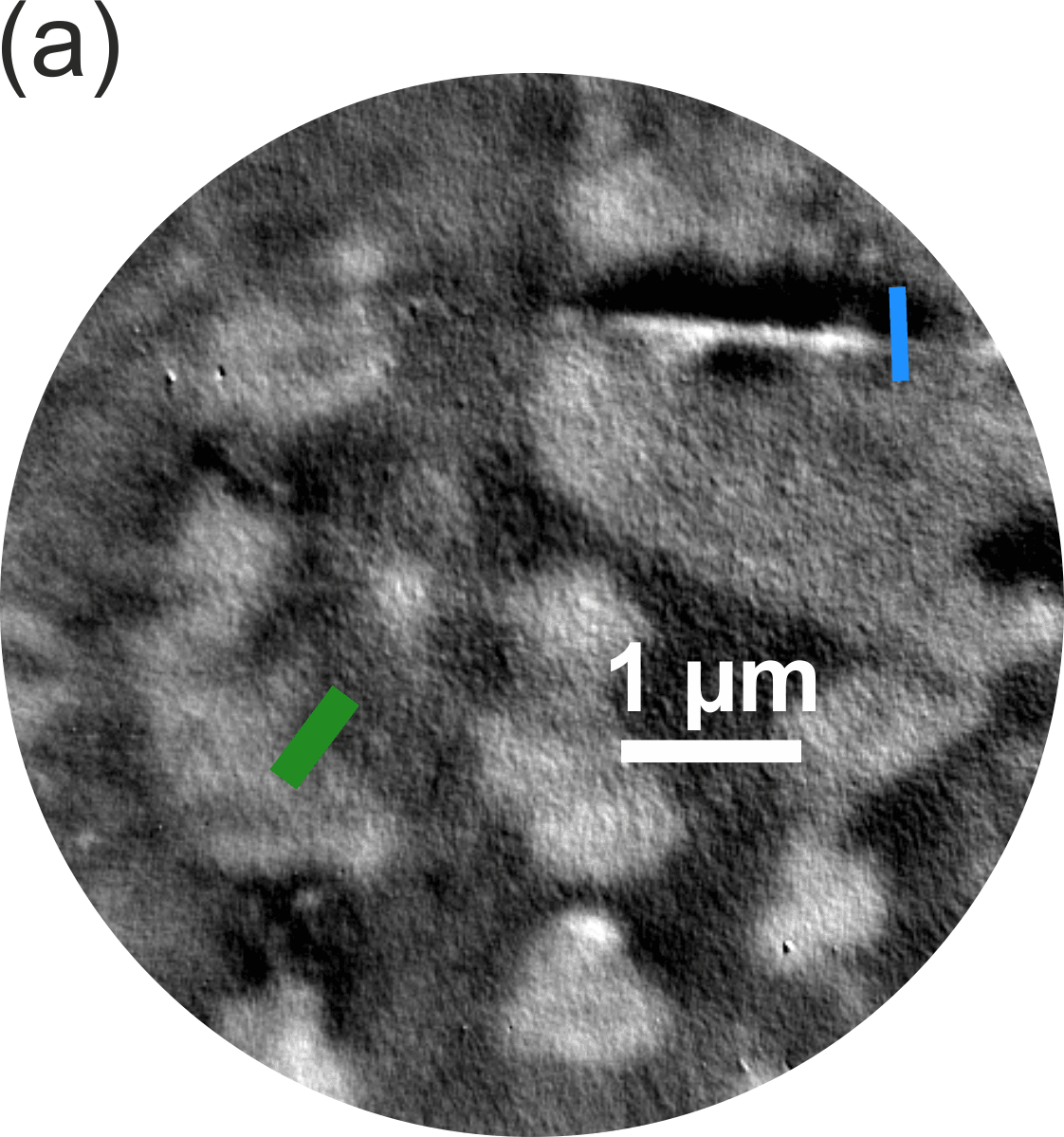}
\includegraphics[height=4.25cm]{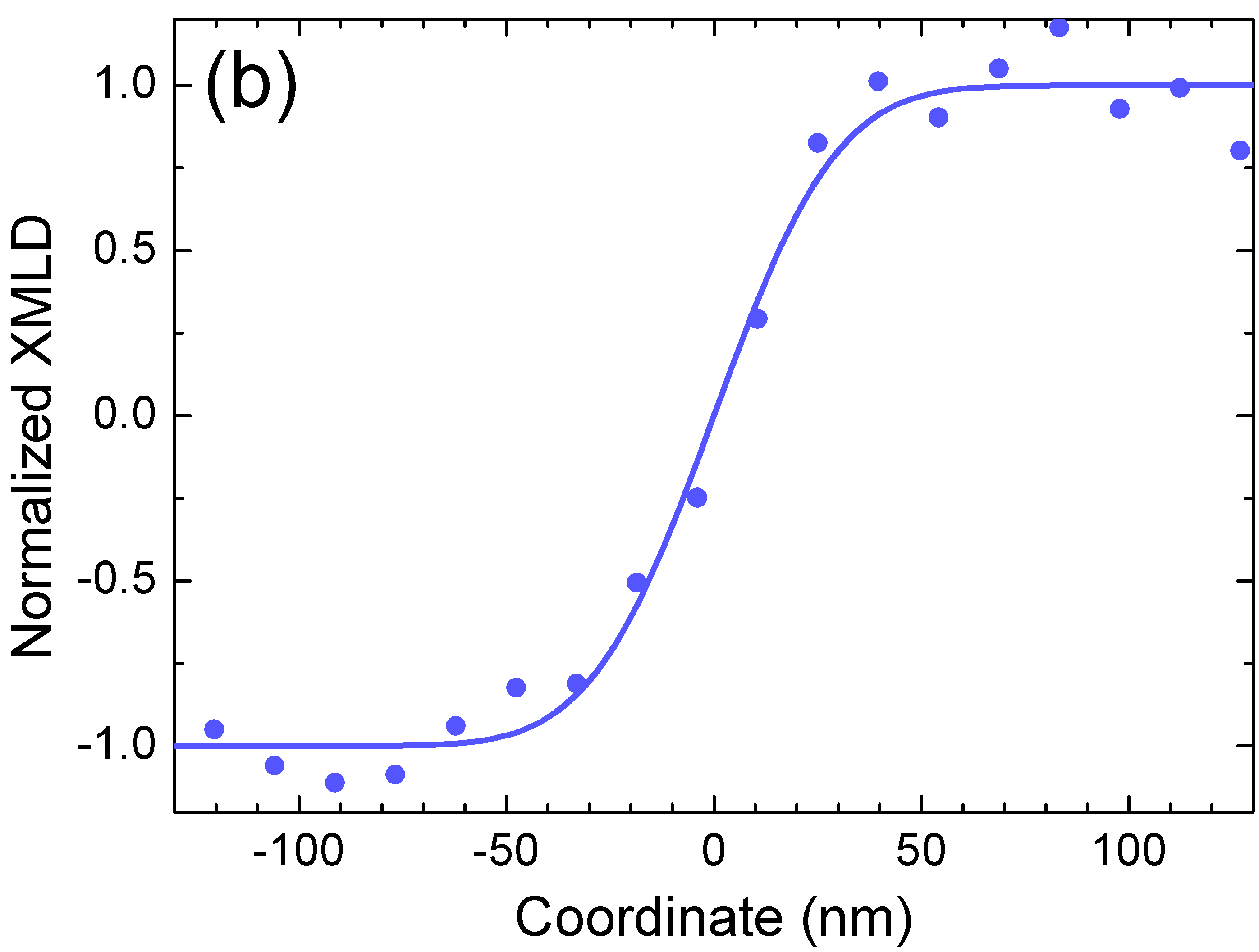}
\includegraphics[height=4.25cm]{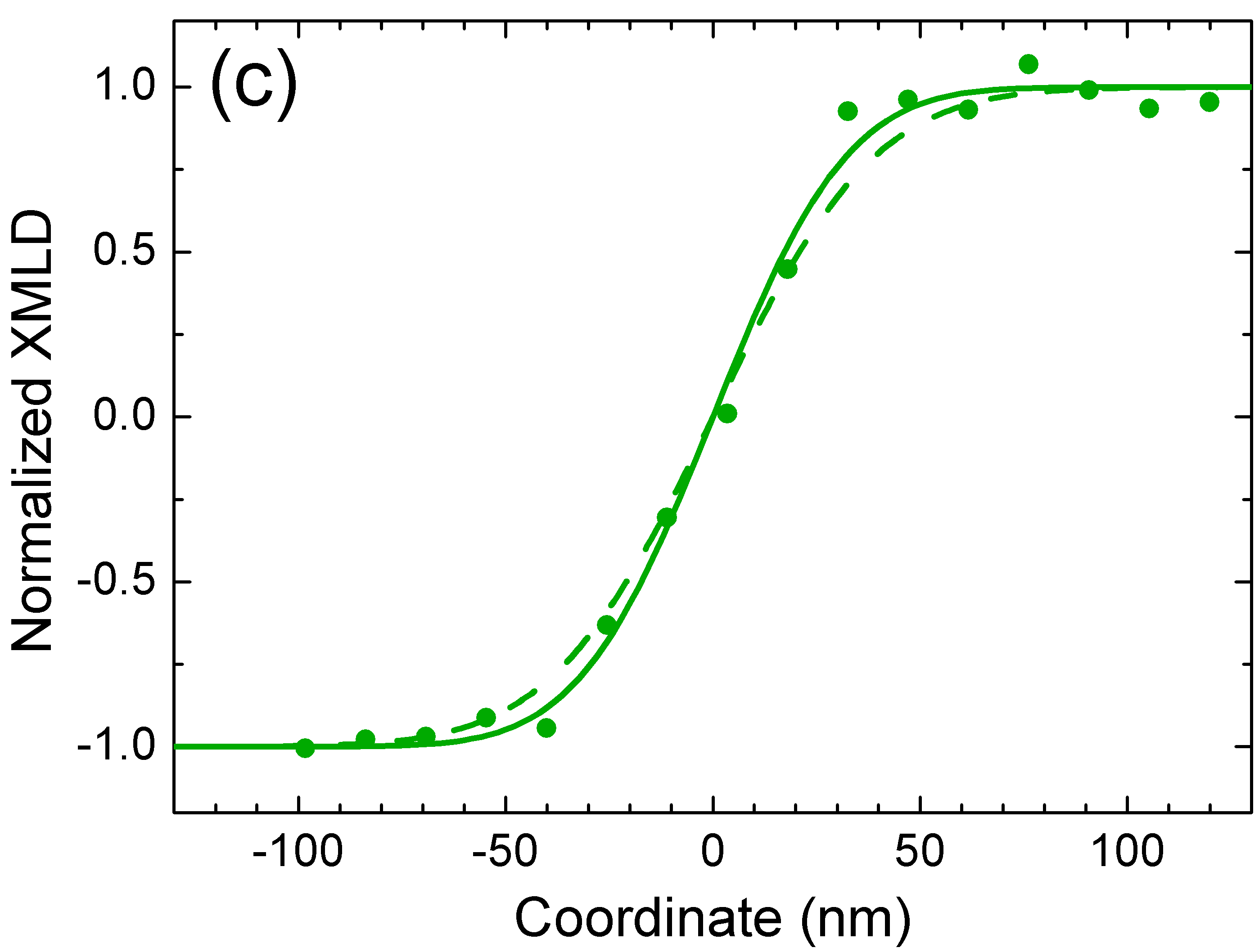}
\caption{(a) Reproduction of Fig.\,\ref{70T_HA} (a) with indicated paths used for measuring the line profiles. (b) Line profile across the topographical structure indicated by the blue line in (a) and corresponding fit with a Gaussian error function. (c) Line profile across the straight domain wall section indicated by the green line in (a) and fits corresponding to $2K_4/J$\,=\,2$\times$10$^{-4}$ (solid line) and $2K_4/J$\,=\,0.5$\times$10$^{-4}$ (dashed line).}
\label{ResProf} 
\end{figure}

However, in an XMLD-PEEM experiment, the DW image is broadened due to a finite instrumental resolution, which can be represented as Gaussian function ($Res_\sigma(x)$), with the parameter $2\sigma$ defining the resolution. For the determination of $2\sigma$, an intensity profile was measured across the edge of the defect in the top right part of Fig.\,\ref{70T_HA} (a) (blue line in Fig.\,\ref{ResProf} (a)). Each point is averaged over \SI{150}{\nano\metre} perpendicular to the line. The profile was fitted by the Gaussian error function (Fig.\,\ref{ResProf} (b)), which is a convolution of the step-function and the Gaussian function. We obtain $2\sigma\,\simeq$\,\SI{47}{\nano\metre}.

A domain wall profile was determined across a straight section of a domain wall (green line in Fig.\,\ref{ResProf} (a)) with every point averaged over \SI{300}{\nano\metre} perpendicular to the line, as depicted in Fig.\,\ref{ResProf} (c). The profile was fitted by a convolution of the instrumental resolution function and the determined domain wall profile  $\left(I_{DW}*Res_\sigma\right)(x)$ with the fit parameter $2K_4/J$. The value of $2K_4/J$ providing the best fit is 2$\times$10$^{-4}$. Additionally, we estimated a lower limit of this parameter of 0.5$\times$10$^{-4}$ (see Fig.\,\ref{ResProf} (c)). Based on Eq. (\ref{DWwidth}), this value corresponds to an upper limit for the DW width of \SI{80}{\nano\metre}, which is of the same order of magnitude as the instrumental resolution. Please note that our analysis relies on the assumption of a perfect straight domain wall section. In FM thin films, straight DW sections are favored minimizing stray fields, which are absent in AFM. Therefore, the apparently straight DW section in an AFM might show a variation of the position perpendicular to the profile. Considering the expected DW width, a much higher spatial resolution of better than \SI{10}{\nano\metre} is necessary for detailed investigations of DWs in Mn$_2$Au.

Finally, the DW width provides an additional estimate of the anisotropy constant $K_4$. Using $J$\,=\,\SI{13.5}{\milli\electronvolt} according to Ref. [23] we find $K_4$\,=\,\SI{1}{\micro\electronvolt}/f.u.. This value is the lower boundary for the in-plane anisotropy constant corresponding to an anisotropy field of \SI{0.02}{\tesla} and $H_{SF}$\,=\,\SI{7}{\tesla}.

\section{V. CONCLUSIONS}
Using XMLD-PEEM we obtained images of AFM domains in Mn$_2$Au thin film samples. The easy axis was experimentally determined to be parallel to the crystallographic $\langle$110$\rangle$-directions, in agreement with reports on bulk single crystals. A typical AFM domain size of $\simeq$\SI{1}{\micro\metre} was observed for as grown thin films.

It was possible to manipulate the AFM domains by a large magnetic field of \SI{30}{\tesla} generating a spin-flop transition. From the magnitude of this field we estimate the in-plane magnetic anisotropy constant $K_4$\,$\leq$\, \SI{17}{\micro\electronvolt}/f.u. The samples exposed to a high external field applied along [110]-easy axis show large AFM domains with the Néel vector oriented primarily perpendicular to the field. A strong magnetic field directed along [100]-hard axis results an increase of the domain size preserving almost equal proportion of in-plane Mn$_2$Au domains oriented along the two in-plane easy axes.

A detailed analysis of the measured domain wall profiles indicates a DW width smaller than \SI{80}{\nano\metre}, which is at the limit of the instrumental resolution. Nevertheless, this value can be used to estimate a lower limit of $K_4$ to be \SI{1}{\micro\electronvolt}/f.u.. Thus from the combination of both limits we estimate anisotropy constant $K_4$ to be between 1 and \SI{17}{\micro\electronvolt}/f.u..\\ \\
\begin{acknowledgments}
The research was financially supported by the German Research Foundation (Deutsche Forschungsgemeinschaft) through the Transregional Collaborative Research Center 173 ”Spin+X”, Project A05. A.\,A.\,S. also wishes to acknowledge the fellowship of the MAINZ Graduate School of Excellence. We thank the Diamond Light Source for the allocation of beam time under Proposal No. SI17120-1 and HZB for the allocation of beam time under Proposal No. 17206035-ST. We acknowledge the support of the HLD at HZDR, member of the European Magnetic Field Laboratory (EMFL).
\end{acknowledgments}


\bibliography{DomainsInMn2Au}
\end{document}